\UseRawInputEncoding
\documentclass[aps,prd,10pt,nofootinbib]{revtex4}
\usepackage{graphicx}
\usepackage{amssymb}
\usepackage{hyperref}
\hypersetup{colorlinks=true,linkcolor=blue,citecolor=red,anchorcolor=black}
\begin{document}

\title{Extension of the Goldstone and the Englert-Brout-Higgs mechanisms to non-Hermitian theories}

\author{Philip D. Mannheim}
\affiliation{Department of Physics, University of Connecticut, Storrs, CT 06269, USA \\
 philip.mannheim@uconn.edu\\ }

\date{September 17 2021}

\begin{abstract}
We discuss the extension of the Goldstone and Englert-Brout-Higgs mechanisms to non-Hermitian Hamiltonians that possess an antilinear PT symmetry. We study a model due to Alexandre, Ellis, Millington and Seynaeve and show that for the spontaneous breakdown of a continuous global symmetry we obtain a massless Goldstone boson in all three of the antilinear symmetry realizations:  eigenvalues real, eigenvalues in complex conjugate pairs, and eigenvalues real but eigenvectors incomplete. In this last case we show that  it is possible for the Goldstone boson mode to be a  zero-norm state.  For the breakdown of a continuous local symmetry the gauge boson acquires a non-zero mass by the Englert-Brout-Higgs mechanism in all realizations of the antilinear symmetry, except the one where the Goldstone boson itself has zero norm, in which case, and despite the fact that the continuous local symmetry has been spontaneously broken,  the gauge boson remains massless.

\end{abstract}

\maketitle


\section{\bf ANTILINEAR SYMMETRY}

With the advent of $PT$  theories a new chapter was opened up in  quantum theory. Specifically, Bender and Boettcher \cite{Bender1998} found that the non-Hermitian Hamiltonian $H=p^2+ix^3$ had an energy eigenspectrum that was completely real. While a surprise, it did not actually violate any law of quantum mechanics since while Hermiticity implies reality there is no theorem that says that a non-Hermitian Hamiltonian could not have a completely real energy spectrum. Hermiticity is only \textbf{sufficient} for reality.

\bigskip
All the same, if a non-Hermitian Hamiltonian is nonetheless going to have a completely real eigenspectrum there would have to be some reason for this to be the case. And Bender and Boettcher identified the reason to be that this particular Hamiltonian had a particular antilinear symmetry, viz. $PT$ ($P$ is parity, $T$ is time reversal), with $p\rightarrow p$, $x\rightarrow -x$ and $i\rightarrow -i$. Gradually, it was realized that antilinearity was the \textbf{necessary} condition for reality, with the \textbf{necessary} and \textbf{sufficient} condition being that  the Hamiltonian have an 
antilinear symmetry and that its eigenstates be eigenstates of the antilinear operator (see e.g. \cite{Mannheim2018} and references therein).

\medskip

But there is more to $PT$.

\section{HOW ANTILINEAR SYMMETRY WORKS}
Consider the eigenvector equation 
\begin{eqnarray}
i\frac{\partial}{\partial t}|\psi(t)\rangle=H|\psi(t)\rangle=E|\psi(t)\rangle.
\label{H1b}
\end{eqnarray}
Replace the parameter $t$ by $-t$ and then multiply by some general antilinear operator $A$:
\begin{eqnarray}
i\frac{\partial}{\partial t}A|\psi(-t)\rangle=AHA^{-1}A|\psi(-t)\rangle=E^*A|\psi(-t)\rangle.
\label{H2b}
\end{eqnarray}
If $H$ has an antilinear symmetry so that $AHA^{-1}=H$, then  
\begin{eqnarray}
HA|\psi(-t)\rangle=E^*A|\psi(-t)\rangle.
\label{H2bb}
\end{eqnarray}

\medskip
\noindent
(1) (Wigner): Energies can be real and have eigenfunctions that obey $A|\psi(-t)\rangle=|\psi(t)\rangle$, 

\medskip
\noindent
(2) or energies can appear in complex conjugate pairs that have conjugate eigenfunctions ($|\psi(t)\rangle \sim \exp(-iEt)$ and $A|\psi(-t)\rangle\sim \exp(-iE^*t)$).

\bigskip
\noindent
As to the converse, suppose we are given that the energy eigenvalues are real or appear in complex  conjugate pairs. In such a case not only would $E$ be an eigenvalue but $E^*$ would be too. Hence, we can set $HA|\psi(-t)\rangle=E^*A|\psi(-t)\rangle$ in (\ref{H2b}), and obtain
\begin{eqnarray}
(AHA^{-1}-H)A|\psi(-t)\rangle=0.
\label{H3b}
\end{eqnarray}
Then if the eigenstates of $H$ are complete, (\ref{H3b}) must hold for every eigenstate, to yield $AHA^{-1}=H$ as an operator identity, with $H$ thus having an antilinear symmetry.

\section{A SIMPLE EXAMPLE}

Consider the $2\times 2$ matrix
\begin{eqnarray}
N=\pmatrix{C+A & iB \cr iB &C-A},\quad [PT,N]=0,
\label{GPT20}
\end{eqnarray}
where $A$, $B$ and $C$ are all real. The matrix $N$ is not Hermitian but does have a $PT$ symmetry if we set $P=\sigma_3$ and $T=K$ where $K$ effects complex conjugation. The eigenvalues of $N$ are given by 
\begin{eqnarray}
\Lambda_{\pm}=C\pm (A^2-B^2)^{1/2},
\label{GPT21}
\end{eqnarray}
and they are \textbf{real} if $A^2\ge B^2$ and in a \textbf{complex conjugate pair} if $A^2<B^2$, this actually being generic to non-Hermitian but $PT$-symmetric systems. 

In addition, if $A=B$  the matrix $N$ only has \textbf{one} right-eigenvector and only \textbf{one} left-eigenvector, despite having two solutions to $|M-\lambda I|=0$ (both with $\lambda=C$), viz.

\begin{eqnarray}
\pmatrix{C+A & iA \cr iA &C-A}\pmatrix{a \cr b}=\pmatrix{Ca+Aa+iAb \cr iAa+Cb-Ab}=C\pmatrix{a \cr b}, \\
\nonumber\\
\pmatrix{e , f}\pmatrix{C+A & iA \cr iA &C-A}=\pmatrix{eC+eA+ifA, ieA+fC-fA}=\pmatrix{eC,fC},
\label{H6}
\end{eqnarray}
 viz.
\begin{eqnarray}
b=ia,\qquad \pmatrix{a \cr b}=\pmatrix{a \cr ia},\qquad f=ie,\qquad\pmatrix{e,f}= \pmatrix{e,ie},\qquad \pmatrix{e,ie}\pmatrix{a \cr ia}=0,
\label{H6c}
\end{eqnarray}
and thus cannot be diagonalized by a similarity transformation. It is thus a non-diagonalizable, Jordan-block matrix. 

This particular Jordan-block situation is a case where the Hamiltonian is manifestly non-diagonalizable and thus manifestly non-Hermitian and yet all eigenvalues are real. Note that the overlap of the left-eigenvector and right-eigenvector in this case is zero.

\section{DIAGONALIZING ONLY IF $A\neq B$}

With
\begin{eqnarray}
N=\pmatrix{C+A & iB \cr iB &C-A}
\label{GPT20a}
\end{eqnarray}
introduce
\begin{eqnarray}
S&=&\frac{1}{2(A^2-B^2)^{1/4}}\pmatrix{(A+B)^{1/2}+(A-B)^{1/2}& i[(A+B)^{1/2}-(A-B)^{1/2}]\cr
 -i[(A+B)^{1/2}-(A-B)^{1/2}]&(A+B)^{1/2}+(A-B)^{1/2} },
\end{eqnarray}
\begin{eqnarray}
 S^{-1}&=&\frac{1}{2(A^2-B^2)^{1/4}}\pmatrix{(A+B)^{1/2}+(A-B)^{1/2}& -i[(A+B)^{1/2}-(A-B)^{1/2}]\cr
 i[(A+B)^{1/2}-(A-B)^{1/2}]&(A+B)^{1/2}+(A-B)^{1/2} },
\end{eqnarray}
\begin{eqnarray}
V&=&\frac{1}{(A^2-B^2)^{1/2}}\pmatrix{A & iB \cr -iB &A},\quad V^{-1}=\frac{1}{(A^2-B^2)^{1/2}}\pmatrix{A & -iB \cr iB &A},
\label{GPT22}
\end{eqnarray}
and they effect
\begin{eqnarray}
&&SNS^{-1}=N^{\prime}=\pmatrix{C+(A^2-B^2)^{1/2} & 0 \cr 0 &C-(A^2-B^2)^{1/2}},
\nonumber\\
&& VNV^{-1}=\pmatrix{C+A & -iB \cr -iB &C-A}=N^{\dagger}.
\label{GPT23}
\end{eqnarray}
Here the right-eigenvectors of $N$ that obey $NR_{\pm}=\Lambda_{\pm}R_{\pm}$ are given by the columns of $S^{-1}$, and the left-eigenvectors of $N$ that obey $L_{\pm}N=\Lambda_{\pm}L_{\pm}$ are given by the rows of $S$. Given the right-eigenvectors one can also construct the left-eigenvectors by using the $V$ operator  when $A^2>B^2$, with the left eigenvectors being constructed as $\langle L_{\pm}|=\langle R_{\pm}|V$.
Thus as long as $A\neq B$ ($S$ becomes undefined at $A=B$) we can diagonalize $N$ and can construct a matrix $V$ that effects the pseudo-Hermiticity condition $VNV^{-1}=N^{\dagger}$ that generalizes $H=H^{\dagger}$, just as is characteristic of a matrix with an antilinear symmetry (\cite{Mannheim2018} and references therein). However, with $A=B$ the matrix $N$ becomes of non-diagonalizable  Jordan-block form.

\section{SO WHAT HAPPENS TO THE GOLDSTONE AND ENGLERT-BROUT-HIGGS MECHANISMS?}

Thus if we are going to replace Hermiticity by some antilinearity requirement, then what is going to happen to other standard quantum results that rely on Hermiticity. In \cite{Mannheim2016, Mannheim2018} the $CPT$ theorem ($C$ is charge conjugation) was derived without Hermiticity, with it being shown that with only invariance under the complex Lorentz group (the proper Lorentz group) and probability conservation  the Hamiltonian would necessarily be $CPT$ symmetric. Since $CPT$ defaults to $PT$ for non-relativistic systems this puts the quantum-mechanical $PT$ program on a very secure theoretical footing.

So what about the Goldstone theorem?  Does it need Hermiticity? This then was the brief of  Alexandre,  Ellis, Millington and Seynaeve \cite{Alexandre2018} as followed up by Mannheim \cite{Mannheim2019}, and then by Alexandre,  Ellis, Millington and Seynaeve \cite{Alexandre2019} themselves. In these latter two papers the Englert-Brout-Higgs mechanism was explored in a non-Hermitian context. Subsequent follow up may be found in \cite{Fring2020a,Alexandre2020a,Fring2020b,Mandel2020,Alexandre2020b,Ohlsson2020,Chernodub1,Millington2020,Alexandre2020c,Fring2020c,Chernodub2,Arean2020,Mavromatos2020,Fring2020d,Fring2020e,Alexandre2020d,Fring2021,Correa2021,Korchin2021}.

\section{\bf NON-HERMITIAN MODEL WITH A CONTINUOUS GLOBAL SYMMETRY}

The model introduced by Alexandre,  Ellis, Millington and Seynaeve in \cite{Alexandre2018} consists of two complex (i.e. charged) scalar fields $\phi_1(x)$ and $\phi_2(x)$ with action 
\begin{eqnarray}
I(\phi_1,\phi_2,\phi^*_1,\phi^*_2)=\int d^4x\left[\partial_{\mu}\phi^*_1\partial^{\mu}\phi_1+\partial_{\mu}\phi^*_2\partial^{\mu}\phi_2
+m_1^2\phi_1^*\phi_1-m_2^2\phi^*_2\phi_2-\mu^2(\phi^*_1\phi_2-\phi^*_2\phi_1)-\frac{g}{4}(\phi^*_1\phi_1)^2\right],
\label{GPT1}
\end{eqnarray}
with $m_1^2$, $m_2^2$, $\mu^2$ and $g$ all being real, and $g$ being positive. Here the star symbol denotes complex conjugation, and thus Hermitian conjugation since neither of the two scalar fields possesses any internal symmetry index. Since the action is not invariant under complex conjugation,  the action  is not Hermitian. It is however invariant under the following $CPT$ transformation
\begin{eqnarray}
\phi_1(x_{\mu})\rightarrow \phi^*_1(-x_{\mu}),\quad \phi_2(x_{\mu})\rightarrow -\phi^*_2(-x_{\mu}),
\quad \phi^*_1(x_{\mu})\rightarrow \phi_1(-x_{\mu}),\quad \phi^*_2(x_{\mu})\rightarrow -\phi_2(-x_{\mu}),
\label{GPT2}
\end{eqnarray}
and thus has an antilinear symmetry.

\bigskip
\noindent

As written, the action given in (\ref{GPT1}) is invariant under the electric charge transformation 
\begin{eqnarray}
\phi_1\rightarrow e^{i\alpha}\phi_1,\quad \phi^*_1\rightarrow e^{-i\alpha}\phi^*_1,\quad \phi_2\rightarrow e^{i\alpha}\phi_2,\quad \phi^*_2\rightarrow e^{-i\alpha}\phi_2,
\label{GPT3}
\end{eqnarray}
to thus possess a standard Noether current
\begin{eqnarray} 
j_{\mu}=i(\phi^*_1\partial_{\mu} \phi_1-\phi_1\partial_{\mu} \phi^*_1)+i(\phi^*_2\partial_{\mu} \phi_2-\phi_2\partial_{\mu} \phi^*_2)
\label{GPT4}
\end{eqnarray}
that is conserved in solutions to the equations of motion.

\section{EQUATIONS OF MOTION}

To study the dynamics associated with the action given in (\ref{GPT1}) we have found it convenient \cite{Mannheim2019} to work in the component basis
\begin{eqnarray} 
\phi_1=\frac{1}{\sqrt{2}}(\chi_1+i\chi_2),\quad \phi^*_1=\frac{1}{\sqrt{2}}(\chi_1-i\chi_2),\quad \phi_2=\frac{1}{\sqrt{2}}(\psi_1+i\psi_2),\quad \phi^*_2=\frac{1}{\sqrt{2}}(\psi_1-i\psi_2),
\label{GPT5}
\end{eqnarray}
where all four $\chi_1$, $\chi_2$, $\psi_1$,  and $\psi_2$ are Hermitian.

In the $\chi_1$, $\chi_2$, $\psi_1$,  and $\psi_2$ basis the action takes the form:
\begin{eqnarray} 
I(\chi_1,\chi_2,\psi_1,\psi_2)&=&\int d^4x \bigg{[}\frac{1}{2}\partial_{\mu}\chi_1\partial^{\mu}\chi_1+
\frac{1}{2}\partial_{\mu}\chi_2\partial^{\mu}\chi_2+
\frac{1}{2}\partial_{\mu}\psi_1\partial^{\mu}\psi_1+
\frac{1}{2}\partial_{\mu}\psi_2\partial^{\mu}\psi_2+
\frac{1}{2}m_1^2(\chi_1^2+\chi_2^2)
\nonumber\\
&-&
\frac{1}{2}m_2^2(\psi_1^2+\psi_2^2)-
i\mu^2(\chi_1\psi_2-\chi_2\psi_1)
-\frac{g}{16}(\chi_1^2+\chi_2^2)^2
\bigg{]},
\label{GPT6}
\end{eqnarray}
and with the appearance of the factor $i$ in the $\mu^2$-dependent term, the action now has the characteristic form of the non-Hermitian but $PT$ symmetric $p^2+ix^3$ theory.

For this action the Euler-Lagrange equations of motion take the form
\begin{eqnarray}
-\partial_{\mu}\partial^{\mu} \chi_1&=&-m_1^2\chi_1+i\mu^2\psi_2+\frac{g}{4}(\chi_1^3+\chi_1\chi_2^2),
\nonumber\\
-\partial_{\mu}\partial^{\mu} \chi_2&=&-m_1^2\chi_2-i\mu^2\psi_1+\frac{g}{4}(\chi_2^3+\chi_2\chi_1^2),
\nonumber\\
-\partial_{\mu}\partial^{\mu} \psi_1&=&m_2^2\psi_1-i\mu^2\chi_2,
\nonumber\\
-\partial_{\mu}\partial^{\mu} \psi_2&=&m_2^2\psi_2+i\mu^2\chi_1.
\label{GPT7}
\end{eqnarray}
Now we note that none of these equations is actually invariant under complex conjugation, a concern that was raised by Alexandre,  Ellis, Millington and Seynaeve \cite{Alexandre2018}. We shall leave resolution of this point for the moment, while noting now that its resolution will not change any of our conclusions. 
These equations of motion admit of a tree approximation minimum in which the scalar field expectation values obey 
\begin{eqnarray}
&&m^2_2\bar{\psi}_1-i\mu^2\bar{\chi}_2=0,\quad m_2^2\bar{\psi}_2+i\mu^2\bar{\chi}_1=0,
\nonumber\\
&&m_1^2\bar{\chi}_1-\frac{\mu^4}{m_2^2}\bar{\chi}_1-\frac{g}{4}\bar{\chi}_1^3-\frac{g}{4}\bar{\chi}_1\bar{\chi}_2^2=0,
\nonumber\\
&&m_1^2\bar{\chi}_2-\frac{\mu^4}{m_2^2}\bar{\chi}_2-\frac{g}{4}\bar{\chi}_2^3-\frac{g}{4}\bar{\chi}_2\bar{\chi}_1^2=0.
\label{GPT8}
\end{eqnarray}

\section{TREE APPROXIMATION MINIMUM}

Choosing the minimum in which 
\begin{eqnarray}
\frac{g}{4}\bar{\chi}_1^2=m_1^2-\frac{\mu^4}{m_2^2}, \quad \bar{\psi}_2=-\frac{i\mu^2\bar{\chi}_1}{m_2^2}, \quad \bar{\chi}_2=0, \quad \bar{\psi}_1=0, 
\label{GPT8a}
\end{eqnarray}
and then expanding around this minimum according to 
\begin{eqnarray}
\chi_1=\bar{\chi}_1+\hat{\chi}_1, \quad \chi_2=\hat{\chi}_2, \quad \psi_1=\hat{\psi}_1, \quad \psi_2=\bar{\psi}_2+\hat{\psi}_2
\label{GPT8b}
\end{eqnarray}
yields a first-order term in the equations of motion of the form:

\begin{eqnarray}
\pmatrix{-\partial_{\mu}\partial^{\mu} \hat{\chi}_1 \cr -\partial_{\mu}\partial^{\mu} \hat{\psi}_2 \cr -\partial_{\mu}\partial^{\mu} \hat{\chi}_2 \cr -\partial_{\mu}\partial^{\mu} \hat{\psi}_1 }=
\pmatrix{ 2m_1^2-3\mu^4/m_2^2 &  i\mu^2 & 0&0 \cr 
i\mu^2& m_2^2&0&0\cr
0&0&-\mu^4/m_2^2&-i\mu^2\cr
0&0&-i\mu^2&m_2^2}
\pmatrix{ \hat{\chi}_1 \cr  \hat{\psi}_2 \cr  \hat{\chi}_2 \cr  \hat{\psi}_1 }= M\pmatrix{ \hat{\chi}_1 \cr  \hat{\psi}_2 \cr  \hat{\chi}_2 \cr  \hat{\psi}_1 }.
\label{GPT9}
\end{eqnarray}
As we see, with our choice of basis, we have already block-diagonalized the mass matrix $M$, and so we can use our previous $2\times 2$ matrix analysis,  with the lower right $2\times 2$ block having $C+A=-\mu^4/m_2^2$, $C-A=m_2^2$, $B=-\mu^2$. We can readily determine the mass eigenvalues, and obtain
\begin{eqnarray}
|M-\lambda I|=\lambda (\lambda +\mu^4/m_2^2-m_2^2)\left[\lambda^2 -\lambda (2m_1^2+m_2^2-3\mu^4/m_2^2)+2m_1^2m_2^2-2\mu^4\right].
\label{GPT10}
\end{eqnarray}
The mass eigenvalue solutions to $|M-\lambda I|=0$  are thus 
\begin{eqnarray}
\lambda_0=0,\quad \lambda_1&=&\frac{m_2^4-\mu^4}{m_2^2},
\nonumber\\
\lambda_{\pm}&=&\frac{2m_1^2m_2^2+m_2^4- 3\mu^4}{2m_2^2} \pm
\frac{1}{2m_2^2}\left[(2m_1^2m_2^2+m_2^4-3\mu^4)^2+8\mu^4m_2^4-8m_1^2m_2^6\right]^{1/2}
\nonumber\\
&=&\frac{2m_1^2m_2^2+m_2^4-3\mu^4}{2m_2^2} \pm
\frac{1}{2m_2^2}\left[(2m_1^2m_2^2-m_2^4-3\mu^4)^2-4\mu^4m_2^4\right]^{1/2}.
\label{GPT11}
\end{eqnarray}
\bigskip
\bigskip
\noindent
We thus see that one eigenvalue is zero. This then is the Goldstone boson.

\bigskip
\noindent
We should add that being massless does not automatically make a mode be a Goldstone boson. Thus if we set $\mu^4=m_1^2m_2^2$ we find that $\lambda_-=0$, to thus give the $4\times 4$ mass matrix a total of two zero eigenvalues, one in the upper left $2\times 2$ sector and one in the lower right $2\times 2$ sector. However, with this choice we find that both $\bar{\chi}_1$ and $\bar{\psi}_2$ vanish in the tree approximation minimum given in (\ref{GPT8a}). But then all four of $\bar{\chi}_1$, $\bar{\psi}_2$, $\bar{\chi}_2$ and $\bar{\psi}_1$ vanish and we are in a normal, non-spontaneously broken vacuum. Since there is no symmetry that would enforce modes to be massless in the normal vacuum this is then just an artifact of the tree approximation, and neither of the two zero eigenvalues  $\lambda_0$ and $\lambda_-$ would remain massless in higher order. [Moreover, for the upper left $2\times 2$ mass matrix given in (\ref{GPT9}) we have $A-B=m_1^2-3\mu^4/2m_2^2-m_2^2/2-\mu^2$, and if we set $\mu^2=m_1m_2$ (i.e., taking both $m_1$ and $m_2$ to be positive) this becomes $A-B=-(m_1+m_2)^2/2$, with the upper left $2\times 2$ matrix not being Jordan block for any values of the positive $m_1$ and $m_2$.]

\bigskip
\noindent
However, if we do not set $\mu^4=m_1^2m_2^2$ then we would have a bona fide spontaneous breakdown since neither $\bar{\chi}_1$ nor $\bar{\psi}_2$ would then vanish. In that case while $\lambda_-$ would become nonzero, $\lambda_0$ would stay massless, and would now indeed be a Goldstone boson, and would remain so in higher order. Now in the following we shall set $\mu^4=m_2^4$. This would then give us a tree approximation minimum in which both $\lambda_0$ and $\lambda_1$ would vanish, and we discuss this case in detail below. And if we in addition set $m_1^2=m_2^2$ so that $\mu^4=m_1^2m_2^2$ we would even have $\lambda_-=0$, and thus with $\mu^2=m_1^2=m_2^2$ we have no less than three massless modes. But then we would be back to the normal vacuum and none of these modes would be a Goldstone boson and all would acquire masses in higher order. Thus in order to obtain spontaneous breakdown we must not impose $\mu^4=m_1^2m_2^2$. And thus if we want to impose $\mu^4=m_2^4$ we must not also impose $m_1^2=m_2^2$.

\bigskip
\noindent
The author is indebted to Dr. A. Fring for raising the issue of what happens if we set $\lambda_-$ to zero, and for informing him that similar concerns and an even richer structure may be obtained in a generalization of the two-field model to three or more scalar fields \cite{Fring2020a}.

\section{ZERO-NORM GOLDSTONE BOSON}

Originally, Alexandre, Ellis, Millington and Seynaeve only showed the emergence of a Goldstone boson  in  the real eigenvalue realization. Here we see that it holds even if energies are in complex conjugate pairs. 

\bigskip
\bigskip
\noindent
And more, if $\mu_2^4=m_2^4$ (viz. $A=B$) then $\lambda_0$ and $\lambda_1$ are both zero. But we cannot   have two Goldstone bosons, so we must be missing an eigenvector. So then the Hamiltonian must not be diagonalizable, just as is the case if $A=B$. We thus extend the Goldstone theorem of Goldstone, Salam and  Weinberg \cite{Goldstone1962} to Jordan-block Hamiltonians. 

\bigskip
\bigskip
\noindent
But even more, we even find an exception \cite{Mannheim2019}. Now the Goldstone boson has zero norm, carries no probability,  and is not detectable. Thus spontaneous breakdown of a continuous global symmetry does not in fact require the existence of an observable massless particle, with the Goldstone theorem being evaded since the Hilbert space metric is not positive definite.

\section{SPONTANEOUSLY BROKEN NON-HERMITIAN THEORY WITH A CONTINUOUS LOCAL SYMMETRY}

Now that we have seen that we can consistently implement the Goldstone mechanism in a $CPT$-symmetric, non-Hermitian theory, it is natural to ask whether we can also implement the familiar Englert-Brout-Higgs mechanism (Englert and Brout \cite{Englert1964}, Higgs \cite{Higgs1964a,Higgs1964b}, Guralnik, Hagen and Kibble \cite{Guralnik1964}). To this end we introduce a local gauge invariance and a gauge field $A_{\mu}$, and with $F_{\mu\nu}=\partial_{\mu}A_{\nu}-\partial_{\nu}A_{\mu}$ replace (\ref{GPT1}) and (\ref{GPT3})  by
\begin{eqnarray}
I(\phi_1,\phi_2,\phi^*_1,\phi^*_2,A_{\mu})&=&\int d^4x\bigg{[}(-i\partial_{\mu}+eA_{\mu})\phi^*_1(i\partial^{\mu}+eA^{\mu})\phi_1+(-i\partial_{\mu}+eA_{\mu})\phi^*_2(i\partial^{\mu}+eA^{\mu})\phi_2
\nonumber\\
&+&m_1^2\phi_1^*\phi_1-m_2^2\phi^*_2\phi_2-\mu^2(\phi^*_1\phi_2-\phi^*_2\phi_1)-\frac{g}{4}(\phi^*_1\phi_1)^2 -\frac{1}{4}F_{\mu\nu}F^{\mu\nu}\bigg{]},
\label{GPT62}
\end{eqnarray}
and 
\begin{eqnarray}
\phi_1\rightarrow e^{i\alpha(x)}\phi_1,\quad \phi^*_1\rightarrow e^{-i\alpha(x)}\phi^*_1,\quad \phi_2\rightarrow e^{i\alpha(x)}\phi_2,\quad \phi^*_2\rightarrow e^{-i\alpha(x)}\phi_2,\quad
eA_{\mu}\rightarrow eA_{\mu}+\partial_{\mu}\alpha(x).
\label{GPT63}
\end{eqnarray}
With (\ref{GPT2}), the $I(\phi_1,\phi_2,\phi^*_1,\phi^*_2,A_{\mu})$ action is $CPT$ invariant since both  $i$ and $A_{\mu}$ are $CPT$ odd (spin one fields have odd  $CPT$ \cite{Weinberg1995}).

We make the same decomposition of the $\phi_1$ and $\phi_2$ fields as in (\ref{GPT5}), and replace (\ref{GPT6}) by 
\begin{eqnarray} 
I(\chi_1,\chi_2,\psi_1,\psi_2,A_{\mu})&=&\int d^4x \bigg{[}\frac{1}{2}\partial_{\mu}\chi_1\partial^{\mu}\chi_1+
\frac{1}{2}\partial_{\mu}\chi_2\partial^{\mu}\chi_2+
\frac{1}{2}\partial_{\mu}\psi_1\partial^{\mu}\psi_1+
\frac{1}{2}\partial_{\mu}\psi_2\partial^{\mu}\psi_2+
\frac{1}{2}m_1^2(\chi_1^2+\chi_2^2)
\nonumber\\
&-&
\frac{1}{2}m_2^2(\psi_1^2+\psi_2^2)-
i\mu^2(\chi_1\psi_2-\chi_2\psi_1)
-\frac{g}{16}(\chi_1^2+\chi_2^2)^2 
\nonumber\\
&-&eA^{\mu}\left(\chi_1\partial_{\mu}\chi_2-\chi_2\partial_{\mu}\chi_1+
\psi_1\partial_{\mu}\psi_2-\psi_2\partial_{\mu}\psi_1\right)
\nonumber\\
&+&\frac{e^2}{2}A_{\mu}A^{\mu}\left[\chi_1^2+\chi_2^2+\psi_1^2+\psi_2^2\right]-\frac{1}{4}F_{\mu\nu}F^{\mu\nu}
\bigg{]}.
\label{GPT64}
\end{eqnarray}

\section{STILL MASSLESS GAUGE BOSON}

In the tree approximation minimum used above in which  
\begin{eqnarray}
\frac{g}{4}\bar{\chi}_1^2=m_1^2-\frac{\mu^4}{m_2^2}, \quad \bar{\psi}_2=-\frac{i\mu^2\bar{\chi}_1}{m_2^2}, \quad \bar{\chi}_2=0, \quad \bar{\psi}_1=0, 
\label{GPT8bc}
\end{eqnarray}
we induce a mass term for $A_{\mu}$ of the form
\begin{eqnarray}
m^2(A_{\mu})=e^2\left(\bar{\chi}_1^2+\bar{\chi}_2^2+\bar{\psi}_1^2+\bar{\psi}_2^2\right)
=e^2\bar{\chi}_1^2\left(1-\frac{\mu^4}{m_2^4}\right)
=\frac{4e^2}{g}\frac{(m_1^2m_2^2-\mu^4)(m_2^4-\mu^4)}{m_2^6}.
\label{GPT65}
\end{eqnarray}

\bigskip
\noindent
Now something quite startling happens. If we again set $\mu^4=m_2^4$ (viz. $A=B$), we find that the gauge boson \textbf{stays massless} \cite{Mannheim2019}. And yet we have spontaneously broken a continuous symmetry. So what happened to the Englert-Brout-Higgs mechanism? 

\bigskip
\noindent
What happens is that in this limit the Hamiltonian becomes Jordan-block and the  Goldstone boson becomes a zero norm state. The standard proof that the gauge boson must be become massive fails because the proof was only established for a positive norm Hilbert space, where a positive norm Goldstone boson becomes the third component of a then massive gauge boson. 

\section{A PUZZLE, ITS RESOLUTION,  AND ANOTHER PUZZLE}

\bigskip
\noindent
In the global continuous symmetry model we obtained equations of motion of the form
\begin{eqnarray}
-\partial_{\mu}\partial^{\mu} \chi_1&=&-m_1^2\chi_1+i\mu^2\psi_2+\frac{g}{4}(\chi_1^3+\chi_1\chi_2^2),
\nonumber\\
-\partial_{\mu}\partial^{\mu} \chi_2&=&-m_1^2\chi_2-i\mu^2\psi_1+\frac{g}{4}(\chi_2^3+\chi_2\chi_1^2),
\nonumber\\
-\partial_{\mu}\partial^{\mu} \psi_1&=&m_2^2\psi_1-i\mu^2\chi_2,
\nonumber\\
-\partial_{\mu}\partial^{\mu} \psi_2&=&m_2^2\psi_2+i\mu^2\chi_1,
\label{GPT7a}
\end{eqnarray}
However, none of these equations is actually invariant under complex conjugation.  Three resolutions of this issue have been presented in the literature. In their paper Alexandre,  Ellis, Millington and Seynaeve \cite{Alexandre2018}  avoided this problem by changing the variational procedure that led to these equations of motion. Specifically, to the original action they added on a surface term, and the constraint that it put on the variational procedure then led to a set of equations of motion that then were invariant under complex conjugation. Interestingly, this led to the same mass matrix as given above, and a  massless Goldstone boson state was again obtained. 

While doing this is somewhat non-standard for variational procedures (but not completely so since a Hawking-Gibbons surface term plays a similar role for variation of  the Einstein-Hilbert action in general relativity), it would nonetheless be more straightforward if we could avoid the introduction of such surface terms, especially since it would prevent drawing the conclusion that to implement the Goldstone theorem in the $PT$ case one might actually have to resort to such a non-conventional variational procedure.

To this end two alternate resolutions of the problem were presented in \cite{Mannheim2019}. The most straightforward is simply to reinterpret the meaning of the star symbol and identify a field such as $\phi^*$ as the $CPT$ conjugate of $\phi$. With this definition both the action and the equations of motion are then $CPT$ symmetric. In fact this is in keeping with the whole $PT$ or $CPT$ theory program as it is $PT$ or $CPT$ conjugates that are the central  building blocks of the program and not Hermitian conjugates, with $CPT$ invariance being the guiding principle for constructing physical theories and not Hermiticity \cite{Mannheim2018}. With this change in the meaning of the star symbol not affecting the mass matrix all the results that we found above continue to hold.

The second option identified in \cite{Mannheim2019} involves making a specific similarity transformation.
Specifically, with canonical conjugates $\Pi_1=\partial_t\psi_1$ and $\Pi_2=\partial_t\psi_1$ for $\psi_1$ and $\psi_2$ we introduce
\begin{eqnarray}
S(\psi_1)=\exp\left[\frac{\pi}{2}\int d^3x \Pi_1(\textbf{x},t)\psi_1(\textbf{x},t)\right],\quad 
S(\psi_2)=\exp\left[\frac{\pi}{2}\int d^3x \Pi_2(\textbf{x},t)\psi_2(\textbf{x},t)\right],
\label{GPT40}
\end{eqnarray}
and obtain 
\begin{eqnarray}
S(\psi_1)\psi_1S^{-1}(\psi_1)=-i\psi_1,~~ S(\psi_1)\Pi_1S^{-1}(\psi_1)=i\Pi_1,~~
S(\psi_2)\psi_2S^{-1}(\psi_2)=-i\psi_2,~~ S(\psi_2)\Pi_2S^{-1}(\psi_2)=i\Pi_2.
\label{GPT41}
\end{eqnarray}

Since these transformations preserve the equal-time commutation relations $[\psi_1(\textbf{x},t),\Pi_1(\textbf{y},t)]=i\delta^3(\textbf{x}-\textbf{y})$, $[\psi_2(\textbf{x},t),\Pi_2(\textbf{y},t)]=i\delta^3(\textbf{x}-\textbf{y})$, they are fully permissible transformations that do not modify the content of the field theory. Applying (\ref{GPT41}) to $I(\chi_1,\chi_2,\psi_1,\psi_2)$ we obtain 
\begin{eqnarray}
S(\psi_1)S(\psi_2)I(\chi_1,\chi_2,\psi_1,\psi_2)S^{-1}(\psi_2)S^{-1}(\psi_1)=I^{\prime}(\chi_1,\chi_2,\psi_1,\psi_2),
\label{GPT42}
\end{eqnarray}
where
\begin{eqnarray}
I^{\prime}(\chi_1,\chi_2,\psi_1,\psi_2)&=&\int d^4x \bigg{[}\frac{1}{2}\partial_{\mu}\chi_1\partial^{\mu}\chi_1+
\frac{1}{2}\partial_{\mu}\chi_2\partial^{\mu}\chi_2-
\frac{1}{2}\partial_{\mu}\psi_1\partial^{\mu}\psi_1-
\frac{1}{2}\partial_{\mu}\psi_2\partial^{\mu}\psi_2+
\frac{1}{2}m_1^2(\chi_1^2+\chi_2^2)
\nonumber\\
&+&
\frac{1}{2}m_2^2(\psi_1^2+\psi_2^2)-
\mu^2(\chi_1\psi_2-\chi_2\psi_1)
-\frac{g}{16}(\chi_1^2+\chi_2^2)^2
\bigg{]}.
\label{GPT43}
\end{eqnarray}

Stationary variation with respect to $\chi_1$, $\chi_2$, $\psi_1$, and $\psi_2$ replaces (\ref{GPT7}) by
\begin{eqnarray}
-\partial_{\mu}\partial^{\mu} \chi_1&=&-m_1^2\chi_1+\mu^2\psi_2+\frac{g}{4}(\chi_1^3+\chi_1\chi_2^2),
\nonumber\\
-\partial_{\mu}\partial^{\mu} \chi_2&=&-m_1^2\chi_2-\mu^2\psi_1+\frac{g}{4}(\chi_2^3+\chi_2\chi_1^2),
\nonumber\\
-\partial_{\mu}\partial^{\mu} \psi_1&=&m_2^2\psi_1+\mu^2\chi_2,
\nonumber\\
-\partial_{\mu}\partial^{\mu} \psi_2&=&m_2^2\psi_2-\mu^2\chi_1,
\label{GPT44}
\end{eqnarray}
and now each one of the equations of motion is separately invariant under  complex conjugation. Thus this time we do not need to reinterpret the star as a $CPT$ transformation (though we still should of course). Finally, since all we have done is make a similarity transformation, the eigenvalues of the mass matrix do not change and the previous analysis continues to hold.

However, now we have a new problem. $I^{\prime}(\chi_1,\chi_2,\psi_1,\psi_2)$ is Hermitian since all of its coefficients are real and all of its fields are Hermitian. So have we  just recovered the standard Goldstone theorem after all this effort? No, since appearances can be deceptive.

\section{HERMITICITY CANNOT BE DETERMINED BY INSPECTION}

We know that the mass matrix has all three of the $PT$ theory realizations, dependent on the values of its parameters. Thus we have all eigenvalues real, eigenvalues in complex conjugate pairs, and mass matrix not diagonalizable. And yet the action looks to be Hermitian in all three realizations and  this simply could not be case. The conclusion: \textbf{Hermiticity cannot be determined by inspection. Hermiticity is a property of the full Hamiltonian not a property of the  individual terms in it, and that depends on boundary conditions.}

\bigskip
\noindent
\textbf{The secret of  $PT$}: In $H=p^2+ix^3$ both $p$ and $x$ are Hermitian when they act on their own eigenstates, i.e., on  states for which you can throw away surface terms in an integration by parts. However, what matters is how $p$ and $x$ act on the eigenstates of $H$, and then you may or may not be able to throw surface terms away (and you may need to continue into the complex plane in order to be able to do so). And only if you can throw surface terms away, can you say that Hamiltonian is Hermitian. \textbf{Hermiticity is a property that is determined jointly by operators and states.}

\textbf{However, you can determine antilinearity by inspection, as that does only depend on the individual terms in a Hamiltonian. Antilinearity is determined by the operators alone.} Thus $H=p^2+ix^3$ is $PT$ symmetric regardless of boundary conditions.

\section{PATH INTEGRALS AND $PT$ SYMMETRY}

As noted in \cite{Mannheim2013,Mannheim2018} this distinction  between Hermiticity and antilinearity is seen even more sharply in the path integral formulation of quantum theory. We define a generating functional
\begin{eqnarray}
W[J]=\int D[\phi]e^{i[S_{CL}(\phi)+J\phi]/\hbar},
\label{GPT44a}
\end{eqnarray}
where $S_{CL}(\phi)$ is the classical action as evaluated on each classical path that is integrated over. You can make $W[J]$ be $CPT$ symmetric path by path, but there is no meaning as to whether $W[J]$ might be Hermitian. Only after you have constructed the quantum Green's functions and then found the quantum Hilbert space can you even begin to ascertain whether the quantum Hamiltonian might be Hermitian, and we have seen examples in which it is not. \textbf{Antilinearity thus has primacy over Hermiticity.} Moreover, once the quantum theory is constructed the Green's functions will correspond \cite{Bender2008b,Mannheim2009,Jones2009,Mannheim2013} to quantities such as $\langle \Omega_R^{CPT}|T[\phi(x)\phi(y)]|\Omega_R\rangle$ or $\langle \Omega_L|T[\phi(x)\phi(y)]|\Omega_R\rangle$, where $\langle \Omega_R^{CPT}|$ is the $CPT$ transform of the right vacuum $|\Omega_R\rangle$ and $\langle \Omega_L|$ is the left vacuum.

Moreover, we even need $CPT$ to correctly construct the appropriate $S_{CL}(\phi)$ before we even do the path integration. Consider a complex classical scalar field with Lagrangian $L=\partial_{\mu}\phi^*\partial^{\mu}\phi$ and now couple it  to a classical electromagnetic vector potential. Which of the following do we use?
\begin{eqnarray}
L_1=[\partial_{\mu}-eA_{\mu}]\phi^*[\partial^{\mu}+eA^{\mu}]\phi, \qquad 
L_2=[\partial_{\mu}-ieA_{\mu}]\phi^*[\partial^{\mu}+ieA^{\mu}]\phi.
\label{GPT44b}
\end{eqnarray}
$L_1$ is natural if you consider classical physics to be restricted to real numbers. (It is not, it is restricted to c-numbers, i.e., to numbers that commute and not to eigenvalues of Hermitian operators.) But only $L_2$ would give the correct quantum theory. So why should we use $L_2$? Could we argue that $i\partial_{\mu}$ and $eA_{\mu}$ are both Hermitian, and thus we should use $L_2$. Definitely not, as there is no meaning to Hermiticity at the classical level. But we can use $CPT$ symmetry since at the classical level $i\partial_{\mu}$ and $eA_{\mu}$ are both $CPT$ even. Thus even at the classical level where Hermiticity is not even definable, we still  need antilinear symmetry. Thus again, \textbf{antilinearity has primacy over Hermiticity}.

\section{SYMPLECTIC SYMMETRY IN CLASSICAL MECHANICS}

Now it can turn out that path integral does not actually exist if the integration  measure for the paths is real, but does exist if we continue the measure  into the complex plane \cite{Mannheim2018}. If we do have to make such a continuation in the classical theory then the resulting quantum theory is a non-Hermitian $PT$ theory. 

Thus we ask whether  we can justify such a continuation of classical theories into the complex plane. 
With the quantum-mechanical  commutation relation $[x,p]=i$ being invariant under complex transformations of the form $x \rightarrow e^{i\theta}x$, $p \rightarrow e^{-i\theta}p$, we look to see if there is an analogous  complex invariance structure for Poisson brackets in classical mechanics.

Thus consider \cite{Mannheim2013}  a classical system with $n$ coordinates $q_i$, $n$ momenta $p_i$, and generic Poisson bracket
\begin{equation}
\{u,v\}=\sum\limits_{i=1}^{n}\left(\frac{\partial u}{\partial q_i}\frac{\partial v}{\partial p_i}-\frac{\partial u}{\partial p_i}\frac{\partial v}{\partial q_i}\right).
\label{A17}
\end{equation}
If we introduce a $2n$-dimensional vector $\eta$ and a $2n$-dimensional matrix $J$ defined as
\begin{eqnarray}
\eta=\pmatrix{q_i\cr p_i},~~~~J=\pmatrix{0&I\cr -I&0},
\label{A18}
\end{eqnarray}
where $I$ is an $n$-dimensional unit matrix, we can compactly write the generic Poisson bracket as 
\begin{equation}
\{u,v\}=\widetilde{\frac{\partial u}{\partial \eta}}J\frac{\partial v}{\partial \eta},
\label{A19}
\end{equation}
where the tilde symbol denotes transpose. If we now make a phase space transformation to a new $2n$-dimensional vector $\xi$ according to 
\begin{equation}
M_{ij}=\frac{\partial \xi_i}{\partial \eta_j},\qquad \frac{\partial v}{\partial \eta}=\tilde{M}\frac{\partial v}{\partial \xi},\qquad \widetilde{\frac{\partial u}{\partial \eta}}=\widetilde{\frac{\partial u}{\partial \xi}}M,
\label{A20}
\end{equation}
then in these new coordinates and momenta the Poisson bracket takes the form
\begin{equation}
\{u,v\}=\widetilde{\frac{\partial u}{\partial \xi}}MJ\tilde{M}\frac{\partial v}{\partial \xi}.
\label{A21}
\end{equation}
The Poisson bracket will thus be left invariant for any  $M$ that obeys the symplectic symmetry relation
\begin{equation}
MJ\tilde{M}=J.
\label{A22}
\end{equation}

If we introduce generators $G$ defined according to $M=e^{i\omega G}$, such generators will obey 
\begin{equation}
e^{i\omega G}Je^{i\omega \tilde{G}}=J,~~~~GJ+J\tilde{G}=0.
\label{A23}
\end{equation}
Since the matrix $J$ obeys $J^{-1}=\tilde{J}=-J$, the generators will obey
\begin{equation}
\tilde{G}=-J^{-1}GJ=JGJ.
\label{A24}
\end{equation}
Solutions to $\tilde{G}=JGJ$ can be broken into two classes, symmetric generators that  anticommute with $J$, viz. those that obey
\begin{equation}
\tilde{G}=G,\qquad GJ+JG=0,
\label{A25}
\end{equation}
and antisymmetric generators that commute with $J$, viz. those that obey

\begin{equation}
\tilde{G}=-G,\qquad GJ-JG=0.
\label{A26}
\end{equation}
In $N=2n$ dimensions there are $N(N-1)/2$ symmetric generators and $N$ antisymmetric generators, for a total of $N(N+1)/2$ generators. These $N(N+1)/2$ generators close on the Lie algebra $Sp(N)$, the symplectic group in $2n$ dimensions. With the generic Lie algebra commutation relations being of the form $[G_i,G_j]=i\sum_k f_{ijk}G_k$ with real structure coefficients $f_{ijk}$, one can find representations of the symplectic algebra in which all the generators are pure imaginary. Thus if, as is standard in classical mechanics, one takes all angles $\omega$ to be real, canonical transformations effected by $e^{i\omega G}$ will transform a real $\eta$ into a real $\xi$.

\section{COMPLEX SYMPLECTIC SYMMETRY IN CLASSICAL MECHANICS}

However, since the algebra of the generators makes no reference to angles, invariance of the classical Poisson brackets under canonical transformations will persist even if the $\omega$ are taken to be \textbf{complex}. The Poisson brackets of classical mechanics thus possess a broader class of invariances than those associated with real canonical transformations alone since one can transform a real $\eta$ into a complex $\xi$ and still preserve the Poisson bracket algebra. 

Consequently, with both the classical Poisson brackets and the quantum commutators admitting of complex canonical transformations, for every such complex transformation we are able to construct a canonical quantization with an associated correspondence principle. Namely, for each canonically transformed Poisson bracket we associate a canonically transformed quantum commutator, with each associated set of classical coordinates being the eigenvalues of the associated transformed quantum operators.

Now while classical mechanics contains this broad class of complex symplectic transformations, they ordinarily play no role in physics since they contain no additional information that is not already contained in the real symplectic transformations alone.

However, this would not be the case if we were to encounter some form of discontinuity when we continue into the complex coordinate plane. In path integral quantization these discontinuities would occur if the path integral only existed for  a domain of classical paths that were not real. In a canonical quantization these discontinuities would occur if quantum-mechanical wave functions are convergent in the domain associated with some Stokes wedges and divergent in some other Stokes wedges. (See \cite{Bender2007} for a discussion of the role of Stokes wedges in $PT$ theories.)

If the domain of convergence includes the real coordinate axis we are in conventional Hermitian quantum mechanics, and we can take the classical limit to be based on real numbers. However, if wave functions are only convergent in Stokes wedges that do not include the real axis, we are in a non-Hermitian realization of the theory. Now amongst such general non-Hermitian realizations there will be some that are also $PT$ realizations. We can therefore anticipate that the ones that are $PT$ realizations are those in which the classical domain for which the path integral exists and the quantum-mechanical domain for which wave functions exist are either the same or related in some way.

To conclude this section we note that for the simple case of a 4-dimensional phase space (viz. $n=2$) the 4-dimensional transposition matrix $C$ that  is involved in charge conjugation of Dirac spinors and effects $C^{-1}\gamma_{\mu}C=-\tilde{\gamma}_{\mu}$ can also play a role in symplectic transformations. In order to be able to use a $C$ that is, like $J$, antisymmetric, orthogonal, and composed of real elements alone, we take $C$ to be given by its representation in the Weyl basis of the Dirac gamma matrices, viz.
\begin{eqnarray}
C=\pmatrix{-i\sigma_2&0\cr 0&i\sigma_2}.
\label{A27}
\end{eqnarray}
Then, if we now take $\eta$ to be of the form   
\begin{eqnarray}
\eta=\pmatrix{p_1\cr q_1\cr q_2 \cr p_2\cr},
\label{A28}
\end{eqnarray}
we can write the Poisson bracket as
\begin{equation}
\{u,v\}=\widetilde{\frac{\partial u}{\partial \eta}}C\frac{\partial v}{\partial \eta},
\label{A29}
\end{equation}
with the symplectic condition then being given by
\begin{equation}
MC\tilde{M}=C.
\label{A30}
\end{equation}
Moreover, since (\ref{A30}) would not be affected if we were to replace $C$ by $iC$, we would then have an operator $iC$ whose square is one, to thus be reminiscent of the $PT$ theory $C$ operator discussed in \cite{Bender2007}. In such a case it would be $i\{u,v\}$ that would be defined as the Poisson bracket, and under a canonical quantization the quantum commutator $[\hat{u},\hat{v}]$ would be identified with  $\hbar$ times it.

\section{NOT JUST COMPLEX VARIABLES BUT COMPLEX ANALYSIS}

In classical mechanics  point particles  move on one-dimensional real trajectories. In  the classical mechanics of waves the normals to the wavefronts move on one-dimensional real trajectories. Characteristic of one-dimensional real trajectories is that the derivative at any point on the trajectory does not depend on the direction in which the derivative is taken.

So suppose we now continue trajectories into the complex plane. They will still be one-dimensional and will not become two-dimensional. But now the derivative at any point on the trajectory will depend on the direction in which the derivative is taken. However, there is a way to avoid this, namely to impose the Cauchy-Riemann equations, viz. for the complex function $f(x,y)=u(x,y)+iv(x,y)$ where $u(x,y)$ and $v(x,y)$ are both real we impose
\begin{equation}
\frac{\partial u}{\partial x}=\frac{\partial v}{\partial y},\quad\frac{\partial u}{\partial y}=- \frac{\partial v}{\partial x}.
\label{A30a}
\end{equation}
And now the derivative is uniquely defined and does not depend on the direction in which it is taken, and can thus be taken to be $df(z)/dz=\partial u/\partial x+i\partial v/\partial x$. And moreover the function $f(x,y)$ is now amenable to complex analysis.

Thus the natural generalization of real trajectories is trajectories in the complex plane that obey the Cauchy-Riemann equations. It is in this sense that the classical paths in the path integral  measure are to be continued into the complex plane.

\section{SUMMARY}

The $PT$ option for quantum theory is now well established. It in no way changes quantum mechanics. It simply takes advantage of an option that was there from the beginning but had been overlooked. Since having an antilinear symmetry is necessary for obtaining real eigenvalues (and also probability conservation  \cite{Mannheim2018}), it actually represents the most general possible option for quantum mechanics that is allowable. In fact in \cite{Mannheim2018} it was noted that antilinearity should be used as the guiding principle for quantum theory rather than Hermiticity.

While antilinearity allows for real eigenvalues, it has two other realizations, realizations that are foreign to Hermitian theories. Energies could appear in complex conjugate pairs and Hamiltonians need not be diagonalizable at all. In this latter, Jordan-block, realization, the eigenstates of the Hamiltonian have a norm that is not positive but zero.

Some familiar theorems of quantum theory that are based on Hermiticity can be extended to the non-Hermitian domain, namely, the $CPT$ theorem, the Goldstone theorem, and the Englert-Brout-Higgs mechanism. All three are found to hold in the antilinear case without needing to assume Hermiticity, and are even found to hold even if energies  appear in complex conjugate pairs.

However, in the non-diagonalizable Jordan-block case something new occurs. While the spontaneous breakdown of a continuous \textbf{global} symmetry still leads to a massless Goldstone boson, in the Jordan-block case the Goldstone boson has zero norm. It is thus not observable.

And more: the spontaneous breakdown of a continuous \textbf{local} symmetry still leads to a massive gauge boson if energies are real or in complex conjugate pairs. But in the Jordan-block case the gauge boson stays massless, despite the fact that a continuous local symmetry has been spontaneously broken.

\end{document}